\documentclass{article}

\usepackage{arxiv}
\usepackage[utf8]{inputenc} 
\usepackage[T1]{fontenc}    
\usepackage{hyperref}       
\usepackage{url}            
\usepackage{booktabs}       
\usepackage{amsfonts}       
\usepackage[numbers]{natbib}
\usepackage{amsmath, enumitem}
\usepackage{amssymb}
\usepackage{upgreek}
\usepackage{lineno}
\usepackage{soul}
\usepackage{fixltx2e}

\usepackage{nccmath}
\usepackage{ifthen}
\usepackage{color}
\usepackage{graphicx}
\usepackage{algorithm}
\usepackage{algorithmic}
\usepackage{setspace}
\usepackage{multirow}
\usepackage[T1]{fontenc}

\graphicspath{{Fig/}}

\title{Efficient Bayesian Full Waveform Inversion and Analysis of Prior Hypotheses in 3D}
\date{} 					

\author{
	Xuebin Zhao \\
	School of Geosciences \\
	University of Edinburgh\\
	Edinburgh, Unite Kingdom \\
	\And
	Andrew Curtis \\
	School of Geosciences \\
	University of Edinburgh\\
	Edinburgh, Unite Kingdom \\
}

\begin{document}
	\maketitle

\begin{abstract}
Spatially 3-dimensional seismic full waveform inversion (3D FWI) is a highly nonlinear and computationally demanding inverse problem that constructs 3D subsurface seismic velocity structures using seismic waveform data. To characterise non-uniqueness in the solutions, we demonstrate Bayesian 3D FWI using an efficient method called physically structured variational inference, and apply it to 3D acoustic Bayesian FWI. The results provide reasonable posterior uncertainty estimates, at a computational cost that is only an order of magnitude greater than that of standard, deterministic FWI. Furthermore, we deploy variational prior replacement to calculate Bayesian solutions corresponding to different classes of prior information at low additional cost. The results obtained using prior information that models should be smooth show loop-like high uncertainty structures that are consistent with fully nonlinear inversion results presented previously. These structures disappear when smoothing is not imposed, so we conclude that they may be caused by smoothness constraints in tomographic problems. We further analyse a variety of prior hypotheses by constructing Bayesian L-curves, which reveal the sensitivity of the inversion process to different prior assumptions. To our knowledge, this is the first study that allows such prior hypotheses to be compared in probabilistic 3D FWI at feasible computational cost. This work shows that fully probabilistic 3D FWI can be performed and can be used to test different prior hypotheses, at a cost that may be practical, at least in small problems.
\end{abstract}

\section{Introduction}
Seismic full waveform inversion (FWI) uses both phase and amplitude information from observed waveform data to estimate compatible subsurface seismic velocity maps \cite{virieux2009overview}. FWI is often implemented using deterministic methods through a gradient-based optimisation in which a measure of misfit between the observed and simulated seismograms is minimised iteratively. Given the highly nonlinear nature of the FWI problem, achieving a good initial model is crucial to avoid convergence to a locally optimal solution, and due to the ill-posed nature of most FWI problems, additional regularisation is also required to stabilise every inversion \cite{zhdanov2002geophysical, asnaashari2013regularized, aghamiry2018hybrid}. While this approach provides a single point estimate of a solution, unfortunately it can not readily be extended to estimate accurate uncertainty in the solutions, and so can not be used to assess risks in post imaging decision-making processes robustly \cite{arnold2018interrogation, ely2018assessing, zhao2022interrogating, zhang2021interrogation, siahkoohi2022deep}. 

Bayesian inference solves inverse problems within a probabilistic framework, in which a family of all potential solutions and their uncertainties are described by the so-called \textit{posterior} probability distribution function (pdf). The posterior pdf is calculated using Bayes' rule to update \textit{prior} knowledge about model parameters with new information from observed data. Monte Carlo sampling methods are commonly used to sample the Bayesian posterior pdf, in order to quantify uncertainties. However, applying conventional Monte Carlo methods to Bayesian FWI is computationally expensive due to the many model parameters involved, which incurs the curse of dimensionality \cite{curtis2001prior, sambridge2002monte, scales2005uncertainties}. Efficient sampling techniques have been developed in attempts to address this issue, by either reducing the dimensionality (number of unknown parameters) of the sampling problem \cite{ray2016frequency, tsai2023towards, hu2024efficient, mulder2024accelerating} or using data-model gradient information to enhance the sampling efficiency \cite{gebraad2020bayesian, zhao2021gradient, berti2023computationally, zunino2023hmclab}, or a combination of both \cite{biswas2022transdimensional}. Nevertheless, direct posterior sampling methods such as these remain a computational challenge.

Monte Carlo methods are expensive in part because no structure is imposed a priori on the sampling distribution; sampling algorithms must therefore be sufficiently flexible to adapt to, and explore, any pdf topography to any level detail. Variational inference offers an alternative to Monte Carlo methods to solve Bayesian problems. They search for an estimate of the posterior pdf that has a particular structure imposed \textit{a priori}. This usually comes in the form of a family of pdf's which is expected to contain an acceptable approximation to the posterior pdf. The best estimate of the posterior pdf within that family can then be found by optimisation rather than random sampling. Because the range of possible pdf estimates is infinitely smaller than the range of general pdf's explored by Monte Carlo methods, variational methods can be more efficient and easier to scale to high dimensional problems. Recently, variational methods have been applied to 2D Bayesian FWI, and have been shown to reduce its computational cost while still providing acceptable estimates of the posterior distribution \cite{zhang2021bayesianfwi, bates2022probabilistic, wang2023re, izzatullah2023physics, zhao2024physically, yin2024wise, sun2024enabling, xie2024stochastic, zhang2024bayesian, zhao2025uncertainty}.

This work concerns 3D FWI, which typically includes amongst the most computationally challenging problems in subsurface science. Nevertheless, \citet{zhang20233} applied three variational methods to a synthetic acoustic 3D Bayesian FWI problem, demonstrating the feasibility of finding solutions. \citet{lomas20233d} and \citet{walker2024novel} then applied one of the methods to a field dataset collected with an airgun source and ocean bottom nodes over a salt body in the Gulf of Mexico. While there was no independent test of the solution quality, they estimated that uncertainties were higher around the boundaries of the salt body, as is generally expected in true seismic imaging solutions \cite{galetti2015uncertainty}. \citet{hoffmann2024local} also performed 3D FWI and local uncertainty estimation using a source subsampling strategy and ensemble Kalman filter \cite{evensen1994sequential}. In most of these 3D FWI studies a spatial smoothing operator is employed \textit{a priori} to improve the convergence rate of the inversion. However, the obtained posterior distribution may then underestimate uncertainties, which in part exclude the true solution \cite{zhang20233}. Moreover, the type and strength of regularisation applied is effectively a subjective hypothesis, and is difficult to choose in practice because almost any form of regularity in the true Earth varies spatially, and so in principle each such hypothesis should be analysed and tested. Considering the huge computational cost of 3D FWI, analysing different hypotheses by comparing Bayesian inversion results that follow from each one is usually impractical.

We demonstrate efficient Bayesian 3D FWI and an approach to analyse different prior hypotheses. The novel contributions of this work can be summarised as follows:
\begin{enumerate}
	\item We show Bayesian 3D FWI with substantially improved efficiency compared to previous methods. The improvement is attributed to the use of a recently introduced variational method: physically structured variational inference \cite[PSVI: ][]{zhao2024physically}, which has proven to provide relatively accurate inversion results including uncertainty estimates at greatly reduced computational cost in 2D FWI problems. Our experiment demonstrates that with PSVI, fully nolinear 3D Bayesian FWI can be solved using only an order of magnitude more computation than traditional, deterministic FWI. To further improve the accuracy of the 3D FWI results reported in previous studies, we use a multiscale inversion strategy \cite{bunks1995multiscale} by performing Bayesian FWI in three frequency bands using relatively low, intermediate, then high frequency data in an attempt to reduce cycle skipping. 
	
	\item To analyse, test and potentially select between different prior hypotheses, we apply a variational prior replacement (VPR) methodology \cite{zhao2024variational} to obtain inversion results corresponding to different prior pdf's at negligible additional computational cost. We compare VPR results with those obtained from independent Bayesian inversion, and find that VPR can not provide perfect estimates of posterior uncertainties due to the extreme complexity of this 3D FWI problem. We therefore introduce a method to fine tune the VPR results with a small amount of extra computation, which improves the results significantly. 
	
	\item We use the posterior pdf's obtained using different prior distributions to construct a so-called \textit{Bayesian L-curve}, from which we analyse different prior assumptions, and select one optimal choice among different hypotheses. 
\end{enumerate}

In the next section we summarise the methods deployed in this study. Next, we apply PSVI to 3D FWI and compare the inversion results with those from other variational and Monte Carlo methods, from which we identify two potential problems that affect the inversion accuracy. We then show how these issues can be addressed and present improved inversion results. In addition, we compare inversion results obtained using different prior hypotheses by building a \textit{Bayesian L-curve}. Finally, we discuss implications of this work and draw conclusions.

\section{Methodology}
\subsection{Variational Bayesian Inversion}
Bayesian inference calculates the \textit{posterior} probability distribution function (pdf) of model parameters $\mathbf{m}$ given observed data $\mathbf{d}_{obs}$ using Bayes' rule
\begin{equation}
	p(\mathbf{m}|\mathbf{d}_{obs}) = \dfrac{p(\mathbf{d}_{obs}|\mathbf{m})p(\mathbf{m})}{p(\mathbf{d}_{obs})},
	\label{eq:bayes}
\end{equation}
where $p(\mathbf{m})$ represents \textit{prior} information about $\mathbf{m}$, and $p(\mathbf{d}_{obs}|\mathbf{m})$ is the \textit{likelihood} of observing $\mathbf{d}_{obs}$ given any value of $\mathbf{m}$. Term $p(\mathbf{d}_{obs})$ is a normalisation constant called the \textit{evidence}.

In this work, we focus on variational inference, in which an optimal variational distribution $q(\mathbf{m})$ is selected from a family of predefined probability distributions, to best approximate the true but unknown posterior distribution $p(\mathbf{m}|\mathbf{d}_{obs})$. This is often accomplished by minimising the Kullback-Leibler (KL) divergence (a measure of difference) between the posterior and variational distributions \cite{kullback1951information}, or equivalently maximising the \textit{evidence lower bound} (ELBO) of $\log p(\mathbf{d}_{obs})$ defined as \cite{blei2017variational}:
\begin{equation}
	\text{ELBO}[q(\mathbf{m})] = \mathbb{E}_{q(\mathbf{m})}[\log p(\mathbf{m}, \mathbf{d}_{obs}) - \log q(\mathbf{m})],
	\label{eq:elbo}
\end{equation}
where $\mathbb{E}_{q(\mathbf{m})}[g(\mathbf{m})] = \int_{\mathbf{m}} q(\mathbf{m}) g(\mathbf{m}) d\mathbf{m}$, which calculates the expectation of an function $g(\mathbf{m})$ with respect to probability distribution $q(\mathbf{m})$. Finding the optimal $q(\mathbf{m})$ by maximising equation \ref{eq:elbo} is an optimisation problem with a fully probabilistic result.

\subsection{Physically Structured Variational Inference (PSVI)}
PSVI is an efficient variational method that, in one implementation, employs a transformed Gaussian distribution with a specific covariance structure to best fit the posterior distribution \cite{zhao2024physically}. A Gaussian distribution $\mathcal{N}(\boldsymbol\mu, \boldsymbol\Sigma)$ is characterised by a mean vector $\boldsymbol\mu$, and a covariance matrix $\boldsymbol\Sigma$. To ensure positive semi-definiteness, the covariance matrix is often re-parametrised using a Cholesky factorisation $\boldsymbol\Sigma = \mathbf{L}\mathbf{L^T}$, where $\mathbf{L}$ is a lower triangular matrix. Variational distribution $q(\mathbf{m})$ can then be obtained by applying an invertible transform $f$ to the Gaussian distribution:
\begin{equation}
	\log q(\mathbf{m}) = \log \mathcal{N}(\boldsymbol\mu, \mathbf{L}\mathbf{L^T}) - \log \left|\det [\partial_{\mathbf{m}}f^{-1}(\mathbf{m})]\right|
	\label{eq:transform}
\end{equation}
Term $\det[\cdot]$ calculates the absolute value of the determinant of the Jacobian matrix $\partial_{\mathbf{m}}f^{-1}(\mathbf{m})$, which accounts for the volume change effected by $f$ \cite{rezende2015variational}. This transform converts a Gaussian random variable defined in an unbounded space into the space of physical model parameters $\mathbf{m}$ which is often bounded \cite{kucukelbir2017automatic}.

Modelling a full covariance matrix requires $n(n+1)/2$ hyperparameters to construct $\mathbf{L}$ ($n$ being the number of model parameters - or the dimensionality of an inverse problem) which is infeasible in 3D FWI problems in which $n$ is typically of order greater than $10^5$. On the other hand, modelling only the diagonal elements (the commonly applied, so-called `mean-field approximation') ignores all correlations between model parameters, resulting in a significant underestimation of uncertainties in the posterior distribution \cite{kucukelbir2017automatic, zhang20233}. To avoid these two extremes, PSVI focuses on capturing the most important (dominant) posterior correlations in the model vector $\mathbf{m}$, guided by prior knowledge of the physics controlling typical imaging inverse problems. Specifically, our implementation of PSVI includes posterior correlations between pairs of locations that are in spatial proximity to each other, typically within one dominant wavelength in FWI problems, since wavefield reflection and refraction respond to physical properties of the subsurface averaged spatially over $\sim$1 wavelength. All other parameter correlations are ignored. This structure of posterior correlations has been observed to emerge naturally in a number of previous nonlinear studies, validating this approach \cite{zhao2024physically}.

We implement PSVI by defining specific off-diagonal elements of $\mathbf{L}$ to be real number parameters to be optimised during inversion; all other elements are set to be zero, thus assuming spatial independence of the corresponding parameter pairs \cite{zhao2024physically}. This results in a sparse structure for $\mathbf{L}$, and thus also the covariance matrix $\boldsymbol\Sigma$, requiring a manageable number of hyperparameters to be optimised in the inversion while still capturing what are typically found to be the most significant correlations. 

We acknowledge that this approach does ignore some inter-parameter correlations which are not strictly zero. However, now that we have managed to perform fully nonlinear Bayesian FWI in previous work \cite[e.g., ][]{zhao2024physically}, we observe clearly that the dominant posterior correlations almost always occur locally, between cells that are up to a wavelength or so apart both horizontally and vertically. This is because within a wavelength there is little ability to resolve much more than the average velocity values, so sub-wavelength velocities can be varied such that they preserve the correct mean value without compromising data fit. Correlations with parameters in cells that are further afield break down very rapidly, because those parameters are also most strongly correlated with values in cells in their local neighbourhoods. In other words, while it is true that there are far field correlations, they are demonstrably (and intuitively) far less important than those in local neighbourhoods. In 3D FWI problems we are obliged to make some approximation to reduce memory requirements when constructing a covariance matrix. To illustrate, in the following section we present a 3D FWI example with a grid size of $101\times101\times63$. If we entertain even only single-precision floating-point format (each element has a size of 4 bytes), storing a full covariance matrix requires about 1 TB memory, which is infeasible for most cases in reality. 

Considering the computational cost of 3D Bayesian FWI, a top priority is to improve its efficiency. PSVI does so at the cost of some loss of generality in the solution found due to the Gaussian foundations for the posterior distribution. \citet{zhao2024physically} demonstrated that the method nevertheless produces reasonable statistical information about the full, nonlinear posterior distribution. We therefore accept the loss of generality and employ the method in this work.

\subsection{Variational Prior Replacement}
The prior replacement was developed for situations in which we wish to calculate various Bayesian solutions to an inverse problem using different classes of prior information given the same observed dataset \cite{walker2014varying}. The variational version of the method used here, called variational prior replacement (VPR), was developed to improve its computational efficiency \cite{zhao2024variational}. Say we have two different prior pdf's $p_{old}(\mathbf{m})$ and $p_{new}(\mathbf{m})$ (hereafter subscripts \textit{old} and \textit{new} are used to denote the order in which the pdf's are estimated or used). According to Bayes' rule (equation \ref{eq:bayes}), the two posterior probability distributions can be calculated by:
\begin{equation}
	p_{old}(\mathbf{m}|\mathbf{d}_{obs}) =  \dfrac{p(\mathbf{d}_{obs}|\mathbf{m})p_{old}(\mathbf{m})}{p_{old}(\mathbf{d}_{obs})} \approx q_{old}(\mathbf{m})
	\label{eq:bayes_vpr1}
\end{equation}
and 
\begin{equation}
	\begin{split}
		p_{new}(\mathbf{m}|\mathbf{d}_{obs}) & = \dfrac{p(\mathbf{d}_{obs}|\mathbf{m})p_{new}(\mathbf{m})}{p_{new}(\mathbf{d}_{obs})} \\
		& = \dfrac{p(\mathbf{d}_{obs}|\mathbf{m})p_{old}(\mathbf{m})}{p_{old}(\mathbf{d}_{obs})} \ \dfrac{p_{new}(\mathbf{m})}{p_{old}(\mathbf{m})} \ \dfrac{p_{old}(\mathbf{d}_{obs})}{p_{new}(\mathbf{d}_{obs})} \\
		& = k \ p_{old}(\mathbf{m}|\mathbf{d}_{obs})\ \dfrac{p_{new}(\mathbf{m})}{p_{old}(\mathbf{m})} \\
		& \approx k\ q_{old}(\mathbf{m}) \ \dfrac{p_{new}(\mathbf{m})}{p_{old}(\mathbf{m})} \\
		& \approx q_{new}(\mathbf{m})
	\end{split}
	\label{eq:bayes_vpr2}
\end{equation}
In equation \ref{eq:bayes_vpr1}, the (old) Bayesian problem given data $\mathbf{d}_{obs}$ and prior pdf $p_{old}(\mathbf{m})$ is solved using variational inference, and the solution $p_{old}(\mathbf{m}|\mathbf{d}_{obs})$ is approximated by the variational distribution $q_{old}(\mathbf{m})$. The third and forth lines in equation \ref{eq:bayes_vpr2} are obtained by substituting equation \ref{eq:bayes_vpr1} into the second line. Term $p_{new}(\mathbf{m})/p_{old}(\mathbf{m})$ updates the solution that includes the old prior to a solution that contains the new one, and we set constant $k = p_{old}(\mathbf{d}_{obs})/p_{new}(\mathbf{d}_{obs})$. This implies that the new posterior distribution can be calculated from the old one by updating prior information \textit{post inversion}, as shown by \citet{walker2014varying}: the inversion corresponding to the new prior distribution does not need to be solved from scratch.

Evaluating $k$ requires the two evidence terms to be calculated which is computationally intractable. \citet{zhao2024variational} introduced a second variational distribution $q_{new}(\mathbf{m})$ to approximate $p_{new}(\mathbf{m}|\mathbf{d}_{obs})$ (fifth line in equation \ref{eq:bayes_vpr2}) by minimising the KL divergence between the new posterior and $q_{new}(\mathbf{m})$, without requiring the value of $k$ to be known explicitly. This step is called \textit{variational prior replacement}.

A variety of variational inference methods can be used to calculate $q_{old}(\mathbf{m})$ and $q_{new}(\mathbf{m})$ \cite{kucukelbir2017automatic, zhang2019seismic, zhao2021bayesian}. In this paper, we employ PSVI because it is efficient, and importantly because its probability value is easy to evaluate for any model parameter values post inversion. Using PSVI to find $q_{old}(\mathbf{m})$ thus allows us to perform VPR efficiently to find $q_{new}(\mathbf{m})$ in the fifth line in equation \ref{eq:bayes_vpr2} that approximates the solution for any of the alternative prior distributions.

\section{3D Bayesian FWI Example}
\begin{figure}
	\centering\includegraphics[width=\textwidth]{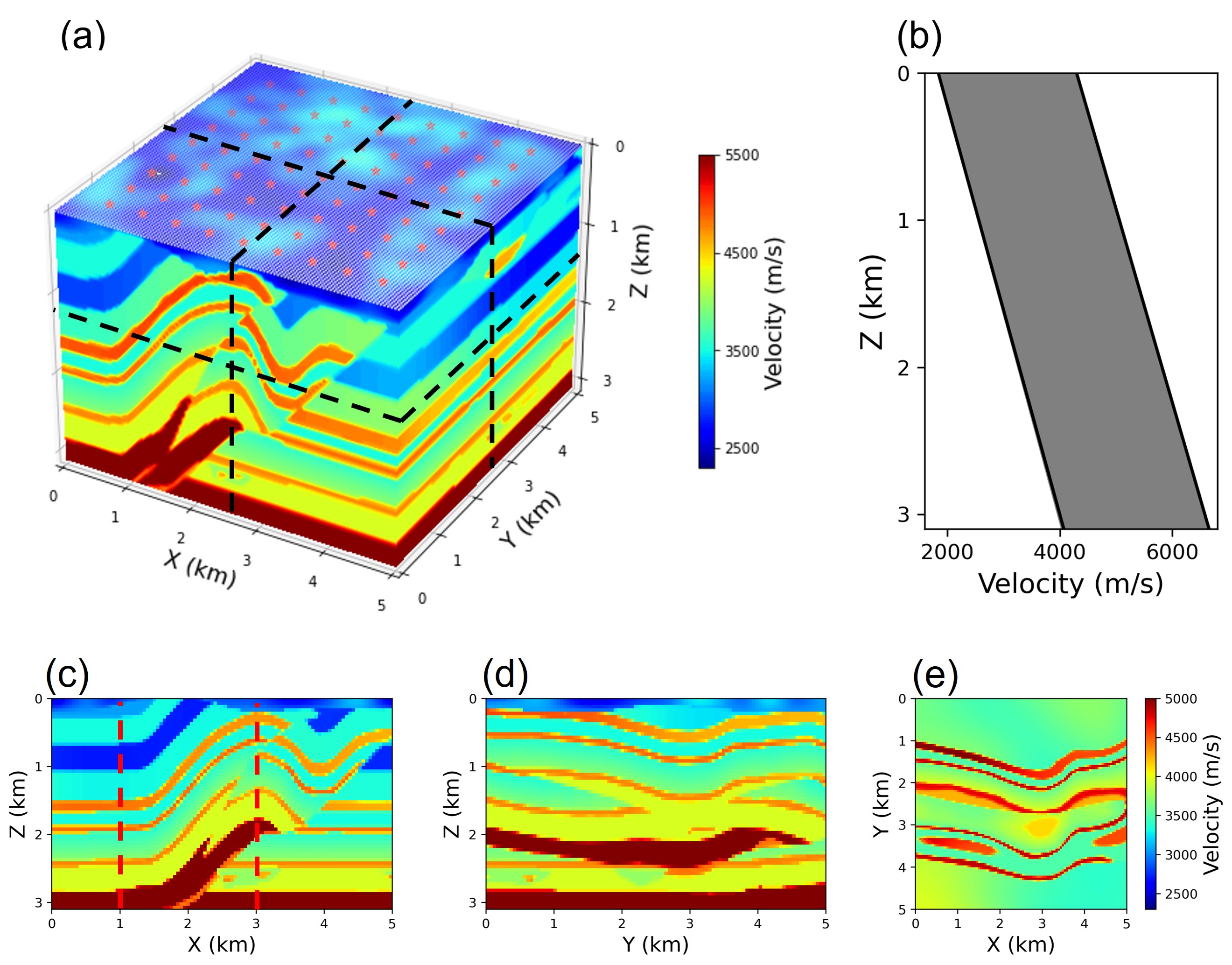}
	\caption{(a) True velocity model. Source and receiver locations are indicated by red stars and white dots at the surface (the latter are so close together that they may be observed as a pale haze). (b) Upper and lower bounds of a uniform prior distribution over seismic velocity at each depth. (c), (d) and (e) Two vertical and one horizontal 2D slices of the true velocity model at Y = 2.5 km, X = 2.5 km, and Z = 1.25 km, respectively. These three sections are marked by dashed black lines in panel (a). Dashed red lines in (c) display locations of two vertical profiles along which posterior marginals are compared in the main text.}
	\label{fig:fwi3d_true_vel_prior}
\end{figure}

We consider a synthetic 3D acoustic FWI problem. As displayed in Figure \ref{fig:fwi3d_true_vel_prior}a, the true velocity model used in this test is a part of the 3D overthrust model \cite{aminzadeh19963}, which contains $63 \times 101 \times 101$ grid cells in Z (depth), Y and X directions, respectively, with a cell size of 50 m in each direction. Figures \ref{fig:fwi3d_true_vel_prior}c -- \ref{fig:fwi3d_true_vel_prior}e show three 2D slices of the true velocity structure at locations Y = 2.5 km, X = 2.5 km, and Z = 1.25 km (horizontal slice), respectively. We deploy 81 impulsive sources (red stars) and 10,201 receivers (white dots) at the surface with spacings of 500 m and 50 m, respectively. Both observed and synthetic data are generated by solving the 3D acoustic wave equation using a time-domain pseudo-analytical method \cite{etgen2009pseudo} with an Ormsby wavelet \cite{ryan1994ricker}. 

We define a uniform prior distribution for the velocity values at different depths, with upper and lower bounds shown in Figure \ref{fig:fwi3d_true_vel_prior}b. We use this uniform prior pdf and a diagonal Gaussian likelihood function (implying uncorrelated data noise - not a requirement of the method) to perform variational Bayesian inversion. 

\subsection{Comparison with Previous Results}
We apply PSVI to the above 3D FWI problem, and compare the inversion results with those obtained from three other methods presented in \citet{zhang20233}: mean field automatic differentiation variational inference \cite[mean field ADVI --][]{kucukelbir2017automatic}, Stein variational gradient descent \cite[SVGD --][]{liu2016stein}, and stochastic SVGD \cite[sSVGD --][]{gallego2018stochastic}. The inversion settings in this example are exactly the same as those used in \citet{zhang20233}, so we are able to compare the four sets of results directly. We set the initial value of PSVI to be a standard Gaussian distribution as suggested by \citet{kucukelbir2017automatic}. This results in a laterally homogeneous initial mean velocity model, with velocity values at different depths being the mean of the uniform prior distribution in Figure \ref{fig:fwi3d_true_vel_prior}b.

\begin{figure}
	\centering\includegraphics[width=\textwidth]{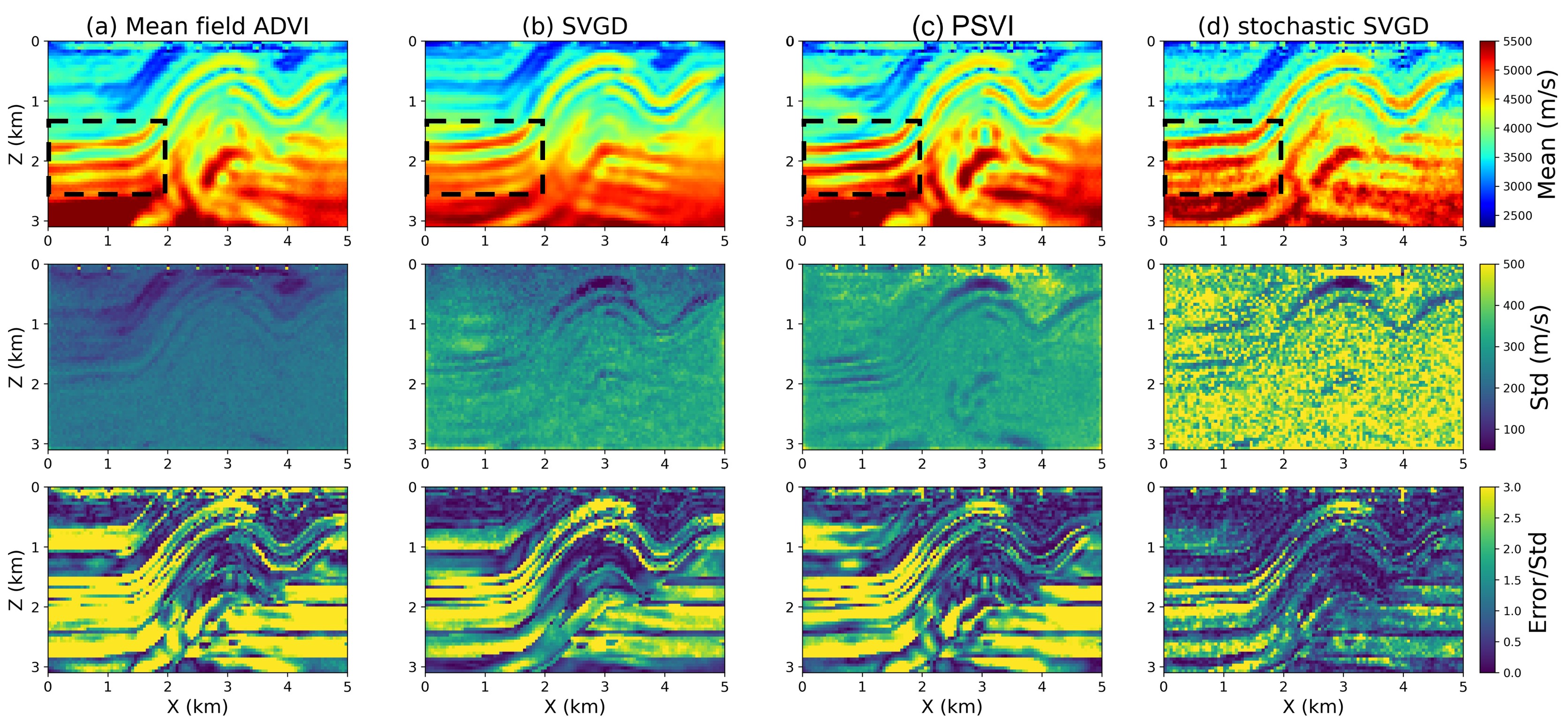}
	\caption{Single frequency band inversion results obtained using 4 different methods, at vertical section Y = 2.5 km. The true velocity model is displayed in Figure \ref{fig:fwi3d_true_vel_prior}c. From top to bottom, each row shows the posterior mean velocity, standard deviation (std), and the relative error maps, respectively, where the relative error is the absolute difference between true and mean velocities (the Error) divided by the standard deviation at each point. Dashed black boxes highlight structures that are not inverted correctly by any method due to cycle skipping.}
	\label{fig:fwi3d_comparison_4methods_meanstd_single_freq}
\end{figure}

Figure \ref{fig:fwi3d_comparison_4methods_meanstd_single_freq} displays the inversion results of the vertical slice shown in Figure \ref{fig:fwi3d_true_vel_prior}c, obtained using different methods indicated in the title of each panel. From top to bottom, each row shows the posterior mean velocity, standard deviation, and the relative error maps, respectively, where the relative error is the absolute difference between the true and mean velocities divided by the standard deviation at each point. Each mean velocity map recovers the main features of the true velocity model. However, we also observe that some structures are inverted incorrectly, such as those inside the dashed black boxes in Figure \ref{fig:fwi3d_comparison_4methods_meanstd_single_freq}, which are supposed to be horizontal according to the true velocity model displayed in Figure \ref{fig:fwi3d_true_vel_prior}c. This is possibly because all four inversions fail to recover correct low-wavenumber components of the true model due to $2\pi$ phase shifts in forward-modelled waveform data fitting the observed data equally well -- a phenomenon often referred to as \textit{cycle skipping} in FWI. Mean field ADVI provides the lowest posterior uncertainties due to its theoretical assumption of an uncorrelated Gaussian posterior pdf. Uncertainty values from SVGD and PSVI are larger than those from mean field ADVI, and are smaller than those from sSVGD. Most of the relative errors in the four plots in the bottom row remain relatively high since they all underestimate the posterior uncertainties to some extent for such a high dimensional inverse problem.

\begin{figure}
	\centering\includegraphics[width=\textwidth]{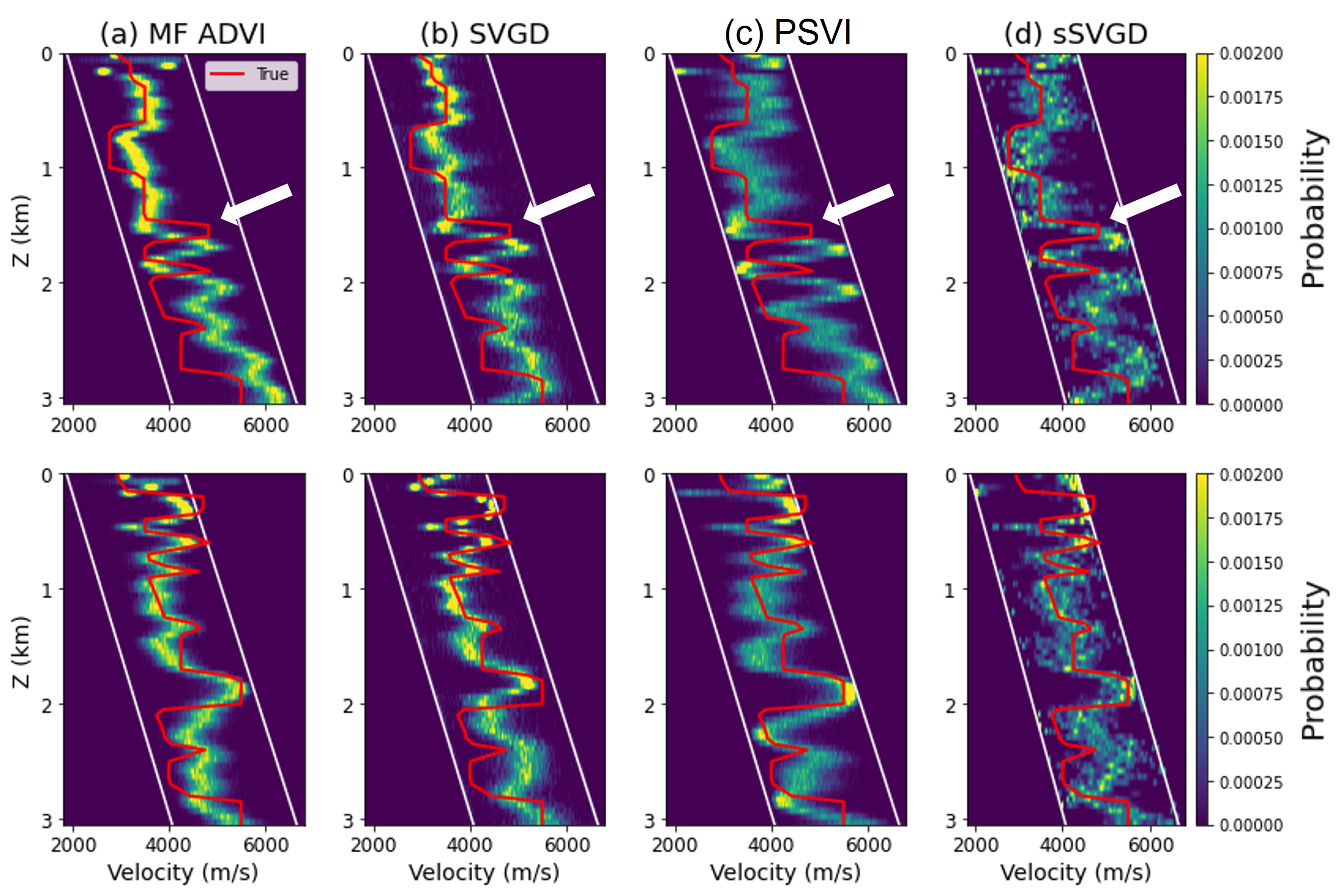}
	\caption{Posterior marginal pdf's obtained using different methods, extracted at two locations marked by red dashed lines in Figure \ref{fig:fwi3d_true_vel_prior}c. Red lines show the true velocity profile. White arrows highlight posterior marginal pdf's that present incorrect phase information caused by cycle skipping.}
	\label{fig:fwi3d_comparison_4methods_marginals_single_freq}
\end{figure}

To justify the above statements and to better compare the 4 sets of results, in Figure \ref{fig:fwi3d_comparison_4methods_marginals_single_freq} we compare the posterior marginal distributions obtained from the 4 methods along two vertical profiles at horizontal locations of 1 km (top row) and 3 km (bottom row), respectively. The locations of these two profiles are displayed by dashed red lines in Figure \ref{fig:fwi3d_true_vel_prior}c. In Figure \ref{fig:fwi3d_comparison_4methods_marginals_single_freq}, red lines show the true velocity profile. Overall, the marginal distributions in Figures \ref{fig:fwi3d_comparison_4methods_marginals_single_freq}c and \ref{fig:fwi3d_comparison_4methods_marginals_single_freq}d present similar features and are broader than those in Figures \ref{fig:fwi3d_comparison_4methods_marginals_single_freq}a and \ref{fig:fwi3d_comparison_4methods_marginals_single_freq}b, indicating that both PSVI and sSVGD provide better posterior uncertainty estimates on average compared to the other two methods.

However, the above 4 results, especially those obtained using PSVI and sSVGD, show some consistent but incorrect patterns. First, the posterior marginals in the first row in Figure \ref{fig:fwi3d_comparison_4methods_marginals_single_freq} provide biased phase information in which the true velocity profile is excluded from the posterior pdf's, as marked by white arrows. PSVI uncertainties are significantly narrower in that section of the profile accentuating the error, presumably because the algorithm is stuck in a local minimum of the variational optimisation problem that happens to have low values for the velocity standard deviations. This section corresponds to the results highlighted by the dashed black boxes in Figure \ref{fig:fwi3d_comparison_4methods_meanstd_single_freq}, due to the incorrectly inverted low-wavenumber component (cycle skipping). Second, in Figure \ref{fig:fwi3d_comparison_4methods_marginals_single_freq} we observe roughly the same magnitude of posterior uncertainties (i.e., similar width of those marginal pdf's) in both shallower and deeper parts of the model. Given that the sensitivity of surface seismic data should often be higher in the near surface than in deeper parts of the Earth, we would expect to observe higher uncertainties at depth. This is caused by additional prior information (spatial smoothness) imposed into the inversion.

In the following section, we solve these two issues to improve the inversion results.

\subsection{Improved Inversion Results}
In this section, we show how to improve the 3D FWI results. To avoid cycle skipping, similarly to the common strategy in linearised FWI problems, we employ a multiscale inversion approach by inverting waveform data from low frequency to high frequency \cite{bunks1995multiscale}. We consider three frequency bands referred to as the low (1 -- 6 Hz), intermediate (1 -- 10 Hz), and high frequency band (1 -- 13 Hz). Variational optimizations using data in intermediate and high frequency bands used results from the previous band as a starting point. The starting point for the low frequency band variational inversion is set to be a standard Gaussian distribution within a unconstrained space, as shown in equation \ref{eq:transform} and suggested by \cite{kucukelbir2017automatic}. In preparation for subsequent sections, we refer to the final solution using high frequency data as $q_{old}(\mathbf{m})$.

Any regularisation (a form of prior information) is somewhat arbitrary as we will never know whether the imposed prior information is consistent with the true Earth or not \textit{a priori}, and thus whether the regularisation will spuriously bias the inversion results or not. We therefore remove all regularisations (note that spatial smoothing is imposed in the previous tests), and simply use the uniform pdf displayed in Figure \ref{fig:fwi3d_true_vel_prior}b as prior information for FWI. As long as the posterior solution is not unduly constrained by the uniform distribution bounds, we assume that the inversion is constrained mainly by the likelihood function and hence by observed waveform data. Note that this will increase the effective support of the posterior pdf (the hyper-volume over which it is significantly greater than zero), and thus will also increase the complexity of this FWI problem, compared to more strongly regularised problems (e.g., with tighter uniform bounds). Due to limited computational resources, we only perform PSVI in this test. After obtaining reasonable inversion results using this non-informative, broad prior distribution, in the next section we analyse different prior assumptions efficiently using a variational prior replacement methodology.

\begin{figure}
	\centering\includegraphics[width=\textwidth]{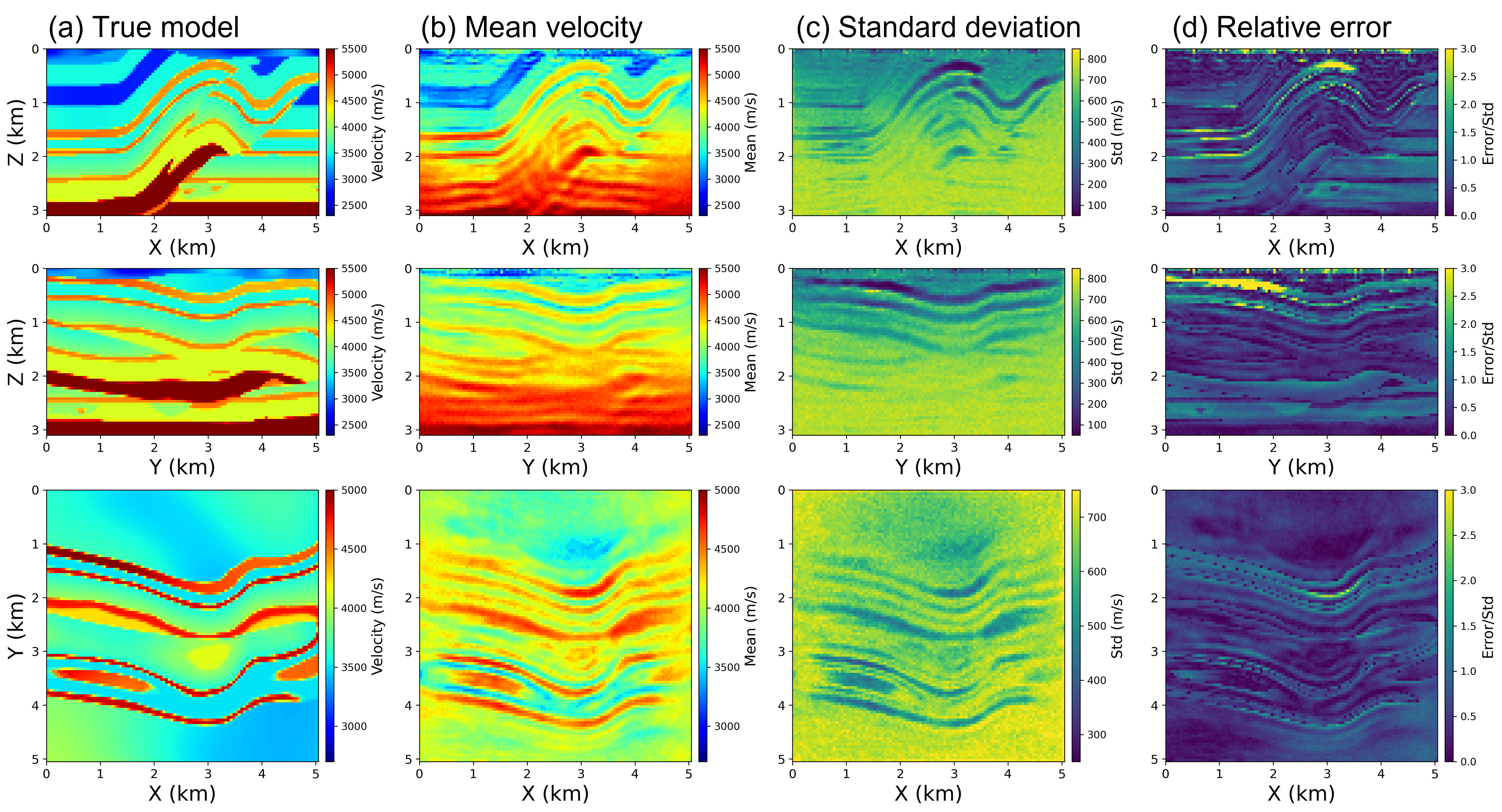}
	\caption{High frequency 3D variational Bayesian FWI results obtained using PSVI and the uniform prior distribution displayed in Figure \ref{fig:fwi3d_true_vel_prior}b. Top two rows show vertical sections at Y = 2.5 km and X = 2.5 km; the bottom row shows a horizontal section at Z = 1.25 km. Panels (a) -- (d) display the true velocity model, the mean, the standard deviation and the relative error (calculated by the absolute difference between the true and mean velocities divided by the standard deviation at each point) maps of the posterior distribution, respectively.}
	\label{fig:fwi3d_highf_3sections_uniform}
\end{figure}

Figure \ref{fig:fwi3d_highf_3sections_uniform} displays the final inversion results obtained using the high frequency band data; interim results using low and intermediate frequency bands are presented in Appendix A. The top two rows display two vertical sections at Y = 2.5 km and X = 2.5 km. The bottom row illustrates one horizontal section at a depth Z = 1.25 km. The three sections are marked by dashed black lines in Figure \ref{fig:fwi3d_true_vel_prior}a. From left to right, Figures \ref{fig:fwi3d_highf_3sections_uniform}a -- \ref{fig:fwi3d_highf_3sections_uniform}d show the three true velocity sections, the average velocity, the standard deviation, and the relative error maps of the posterior probability distribution, where the relative error is the absolute difference between the true and mean velocities divided by the standard deviation at each point.

The three mean sections resemble the true velocity structures (note that even for an entirely correct solution, this need not be the case in probabilistic inversions since the mean of the models is a statistic, not a model in itself). This is especially true in the shallow subsurface above 2 km where high resolution structures are accurately inverted. However, the mean and true values diverge at deeper levels, possibly due to reduced data sensitivity since the standard deviations broadly increase with depth. The overall relative errors are less than 3, indicating that differences between the true and inverted mean models are within three standard deviations, as would be expected of the true solution to this Bayesian problem. Note that in Figure \ref{fig:fwi3d_highf_3sections_uniform}d, higher relative errors (shown in yellow) are present near the surface because in fact the true model deliberately lies outside of the prior bounds shown in Figure \ref{fig:fwi3d_true_vel_prior}b, to assess the method's behaviour in such circumstances \cite{zhang20233}. Overall, the inverted images and the posterior uncertainties in Figure \ref{fig:fwi3d_highf_3sections_uniform} are significantly more accurate than those displayed in Figure \ref{fig:fwi3d_comparison_4methods_meanstd_single_freq}.

\subsection{Testing Different Prior Hypotheses}
In addition to the uniform prior distribution, we now consider a set of smoothed prior distributions. Define a second-order finite difference operator $\mathbf{S}$ to calculate curvature of model parameter values between adjacent grid cells in a spatial 3D velocity model represented by model vector $\mathbf{m}$ \cite{zhao2024variational}.
This is further applied to $\mathbf{m}$ to impose spatial smoothness, by imposing a Gaussian distribution on the product $\mathbf{Sm}$
\begin{equation}
	p(\mathbf{Sm}) \propto \exp \left(-\frac{(\mathbf{Sm})^T (\mathbf{Sm})}{2 \sigma_s^2}\right),
	\label{eq:sm}
\end{equation}
where $\sigma_s$ is a hyperparameter that controls the strength of spatial smoothness assumed (a smaller $\sigma_s$ causes the probability of any particular rough model to become lower). Equation \ref{eq:sm} can be interpreted as applying a Tikhonov matrix $\mathbf{S}$ to $\mathbf{m}$ \cite{golub1999tikhonov}. Then the smoothed model prior pdf can be expressed as
\begin{equation}
	p_{smooth}(\mathbf{m}) \propto p(\mathbf{Sm})p_{old}(\mathbf{m}),
	\label{eq:smooth_prior}
\end{equation}
where $p_{old}(\mathbf{m})$ is the (old) uniform prior distribution defined in the previous section. We consider 7 smoothed prior pdf's with different $\sigma_s$ values: 2000, 1000, 500, 250, 125, 62.5 and 31.25, respectively. In the later part of this paper, we also denote the uniform prior distribution as $\sigma_s = \infty$ indicating no smoothing. These 8 prior pdf's are used to mimic a real situation in which we have different prior hypotheses and wish to analyse their consequences.

\begin{figure}
	\centering\includegraphics[width=\textwidth]{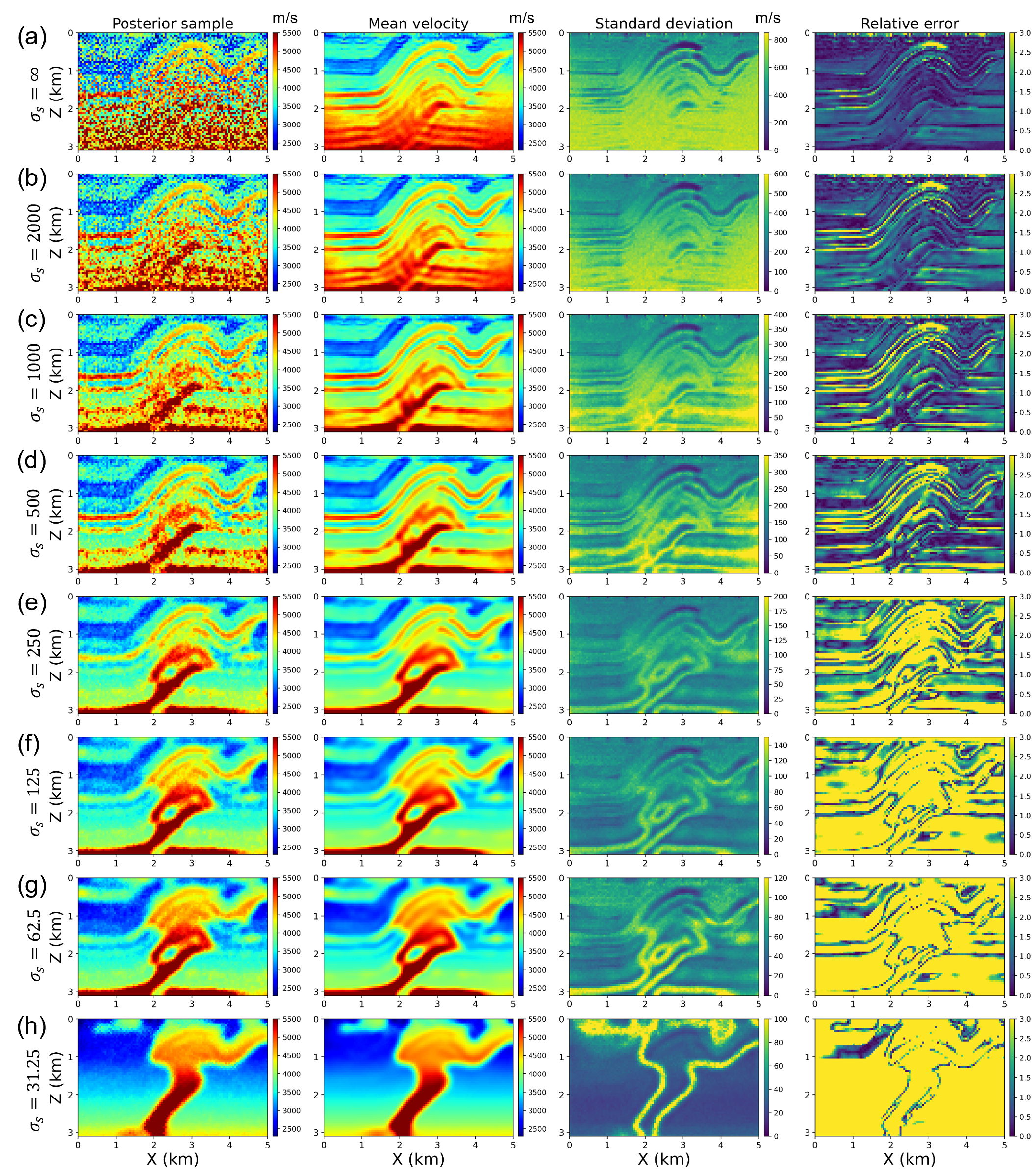}
	\caption{Inversion results obtained using (a) the uniform prior pdf and (b) -- (h) the 7 smoothed priors with decreasing values of $\sigma_s$ (equivalently increasing smoothness), in a vertical section Y = 2.5 km. From left to right, they represent one posterior sample, the mean velocity, the standard deviation and the relative error maps.}
	\label{fig:fwi3d_psvi_diff_smooth_vpr_mean_std_error}
\end{figure}

The posterior pdf's using each of the 7 smoothed priors are obtained by replacing the uniform prior pdf in inversion results $p_{old}(\mathbf{m}|\mathbf{d}_{obs})$ from the previous section (Figure \ref{fig:fwi3d_highf_3sections_uniform}) by the respective smoothed prior pdf's, using variational prior replacement (VPR). Thus, no additional inversion is performed to obtain these results. In Appendix B, we verify the results of VPR by comparing them to those obtained by performing independent Bayesian inversions (verification of VPR in 2D FWI was also conducted in \citet{zhao2024variational}). Figure \ref{fig:fwi3d_psvi_diff_smooth_vpr_mean_std_error} displays the results using all 8 prior pdf's on vertical section Y = 2.5 km. From left to right, each column shows one posterior sample, the mean, the standard deviation, and the relative error maps of the posterior distribution, respectively. From top to bottom, each row presents the posterior pdf obtained using the $\sigma_s$ value indicated on the left of the figure.

From Figures \ref{fig:fwi3d_psvi_diff_smooth_vpr_mean_std_error}a to \ref{fig:fwi3d_psvi_diff_smooth_vpr_mean_std_error}h, as the magnitude of smoothness increases, stronger prior information favouring spatially smooth structures is injected into the inversion results. Consequently, both the posterior samples and the mean velocity maps become smoother with fewer spatial variations. The standard deviations decrease (note that different colorbar scales are used in the standard deviation maps in the third column); this leads to increased relative errors, as (true) large velocity contrasts between neighbouring cells are precluded by the prior information for small $\sigma_s$. 

\begin{figure}
	\centering\includegraphics[width=\textwidth]{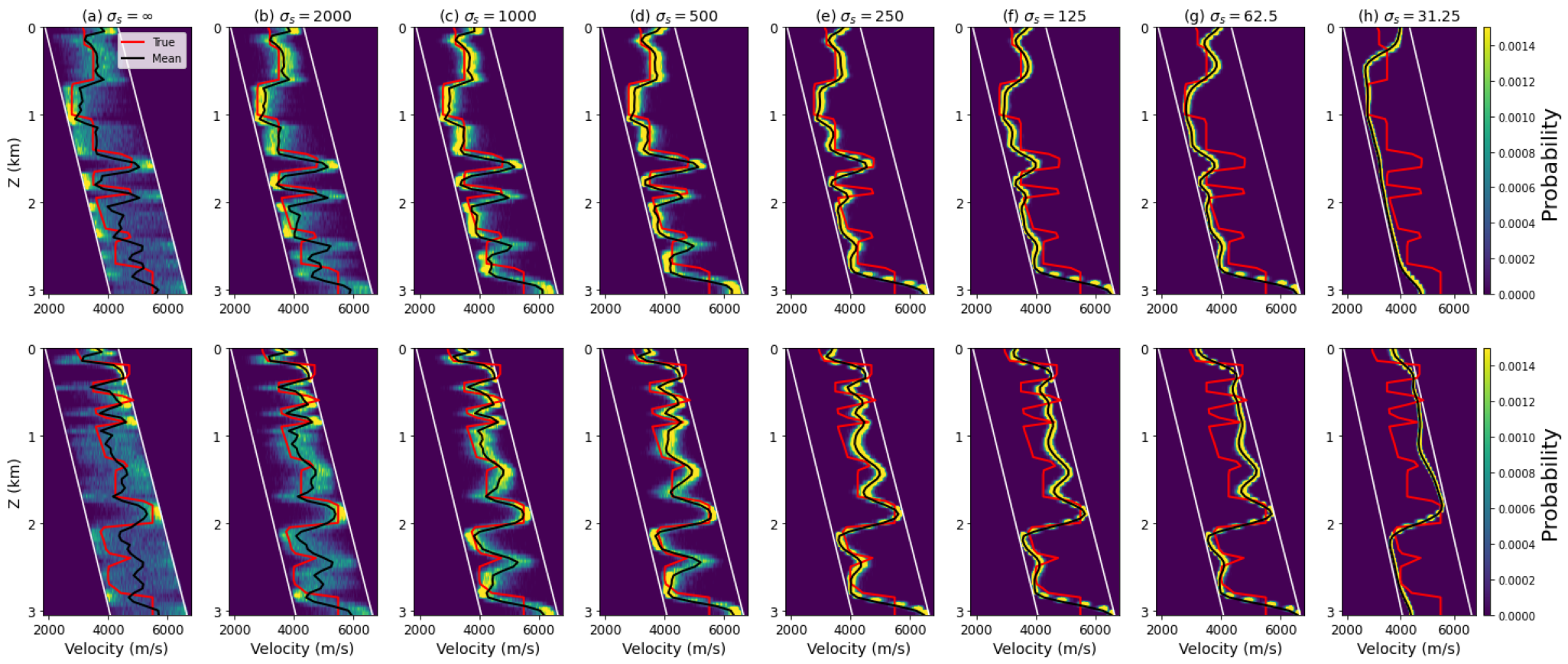}
	\caption{Posterior marginal pdf's at two vertical locations marked by red dashed lines in Figure \ref{fig:fwi3d_true_vel_prior}c. (a) Results obtained using the uniform prior pdf, and (b) -- (h) results using the smoothed prior pdf's with different $\sigma_s$ values.}
	\label{fig:fwi3d_psvi_diff_smooth_vpr_marginals_iy50_ix25_60}
\end{figure}

Figure \ref{fig:fwi3d_psvi_diff_smooth_vpr_marginals_iy50_ix25_60} compares the posterior marginal pdf's of two vertical velocity profiles at locations X = 1km (top row of Figure \ref{fig:fwi3d_psvi_diff_smooth_vpr_marginals_iy50_ix25_60}) and X = 3km (bottom row), marked by red dashed lines in Figure \ref{fig:fwi3d_true_vel_prior}c. Figure \ref{fig:fwi3d_psvi_diff_smooth_vpr_marginals_iy50_ix25_60}a displays the posterior marginal pdf's obtained from the uniform prior distribution, and Figures \ref{fig:fwi3d_psvi_diff_smooth_vpr_marginals_iy50_ix25_60}b -- \ref{fig:fwi3d_psvi_diff_smooth_vpr_marginals_iy50_ix25_60}h show those from the smoothed prior pdf's -- the smoothness values $\sigma_s$ are noted in the title. In each figure, the red line shows the true velocity profile and black line displays the inverted mean velocity profile. We observe that with the increase of the magnitude of smoothness injected by the prior information, the posterior marginal pdf's become smoother with fewer spatial variations (thus lower uncertainties), similarly to the conclusions drawn above.

By comparing each set of results with the true velocity model in Figure \ref{fig:fwi3d_true_vel_prior}c, we find that the posterior pdf's are more tightly constrained around the true values as we increase the magnitude of the smoothness up to $\sigma_s = 500$ in Figure \ref{fig:fwi3d_psvi_diff_smooth_vpr_mean_std_error}d, especially in deeper parts of the model. Interestingly, in the third column in Figures \ref{fig:fwi3d_psvi_diff_smooth_vpr_mean_std_error}c and \ref{fig:fwi3d_psvi_diff_smooth_vpr_mean_std_error}d, we observe higher uncertainties at the boundaries between strata below 1.5 km depth, capturing the expected 'uncertainty-loop' character of posterior uncertainties proposed by \citet{galetti2015uncertainty}. However, Figure \ref{fig:fwi3d_psvi_diff_smooth_vpr_mean_std_error} shows that uncertainty loops appear because smoothness is applied to models, and do not appear when models are allowed to be rough; this is consistent with the results of \citet{galetti2015uncertainty} which were obtained using a trans-dimensional Monte Carlo travel-time tomography method that imposed strict smoothness (indeed, constant velocity) over large areas of model. We therefore qualify the proposition of \citet{galetti2015uncertainty}, by hypothesising that posterior uncertainty loops are in fact a product of smoothness constraints in tomographic problems (including FWI).

The inversion results become worse in Figures \ref{fig:fwi3d_psvi_diff_smooth_vpr_mean_std_error}e -- \ref{fig:fwi3d_psvi_diff_smooth_vpr_mean_std_error}h, as the effect of prior information overpowers information in the observed data, resulting in biased posterior solutions and overly smoothed structures. Not surprisingly, most of the relative errors are then larger than 3. We conclude that the smoothed prior distributions with $\sigma_s = 1000$ or $\sigma_s = 500$ (Figures \ref{fig:fwi3d_psvi_diff_smooth_vpr_mean_std_error}c and \ref{fig:fwi3d_psvi_diff_smooth_vpr_mean_std_error}d) are relatively good choices for this FWI problem.

\begin{figure}
	\centering\includegraphics[width=0.75\textwidth]{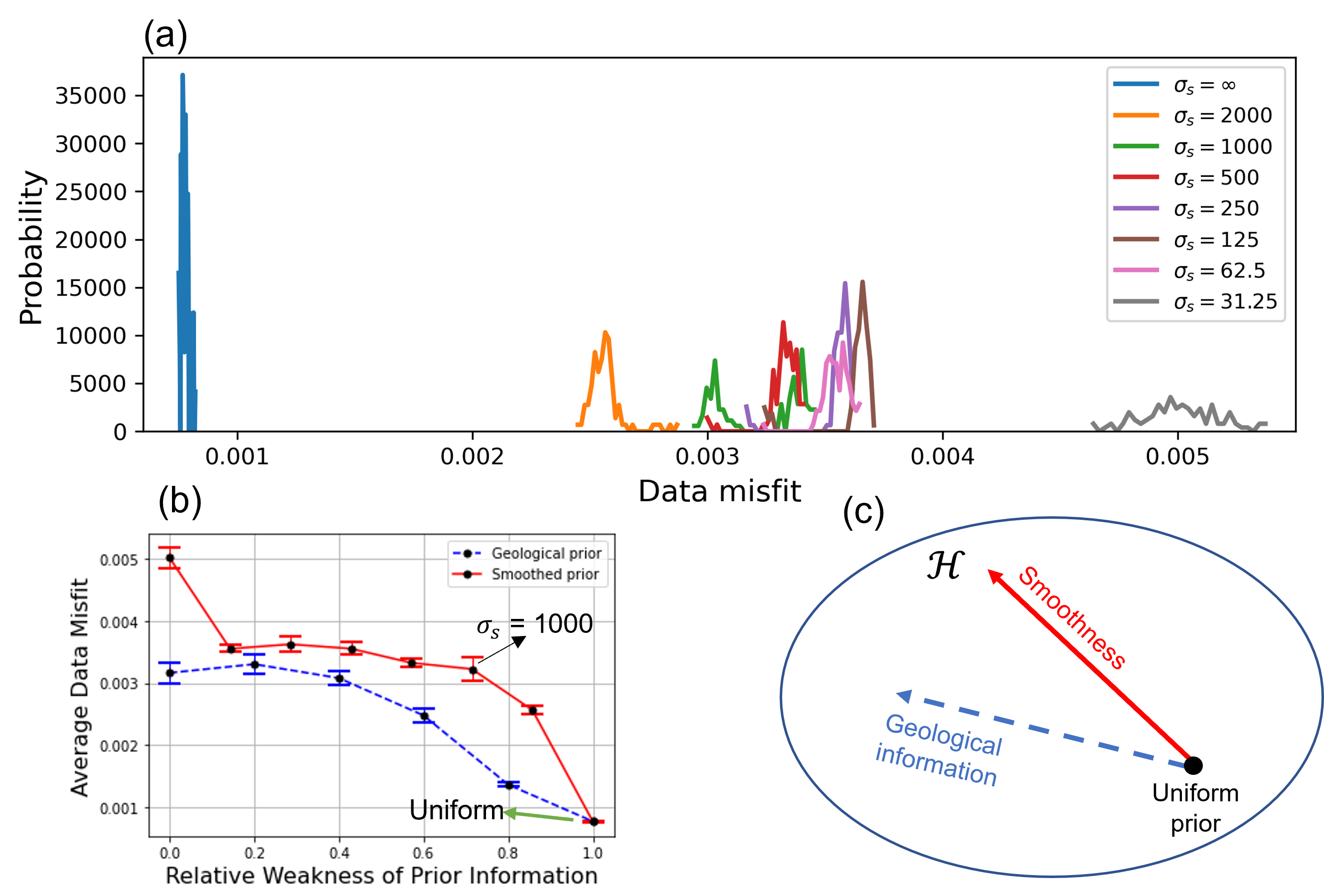}
	\caption{(a) Probability density functions for data misfit values calculated from posterior pdf's displayed in Figure \ref{fig:fwi3d_psvi_diff_smooth_vpr_mean_std_error} using differently smoothed priors. (b) Bayesian L-curves representing the average data misfits and their standard deviations (error bars) with respect to different prior hypotheses. The horizontal axis represents the normalised (relative) weakness of prior information applied by each prior distribution: the value 1 represents the non-informative, uniform prior pdf represented in Figure \ref{fig:fwi3d_true_vel_prior}b. The red line denotes the L-curve found by requiring smoothness in the prior pdf's, and dashed blue line denotes the equivalent curve when requiring consistency with geological prior information. (c) A schematic diagram of the space $\mathcal{H}$ containing all possible prior hypotheses. Black dot represents a non-informative uniform prior distribution, and we denote different prior assumptions such as smoothed and geological prior pdf's in this study by different paths (arrows) through $\mathcal{H}$.}
	\label{fig:fwi3d_misfit}
\end{figure}

However, in reality we never know the true velocity structure, so can not calculate or use relative errors to choose between prior distributions, other than in synthetic tests such as in Figures \ref{fig:fwi3d_psvi_diff_smooth_vpr_mean_std_error} and \ref{fig:fwi3d_psvi_diff_smooth_vpr_marginals_iy50_ix25_60}. In deterministic FWI, an order of preference is often constructed by calculating data misfit values arising from the different inversion results (referred to as an \textit{L-curve}). Following a similar approach, we compare the 8 posterior distributions by drawing N = 100 samples from each posterior pdf and performing forward simulations to calculate corresponding misfit values between the observed and synthetic data. This results in 8 approximate probability distributions over misfit values from the inversion results using different priors, displayed in Figure \ref{fig:fwi3d_misfit}a. As expected, the overall misfit values increase as more smoothness is injected into the inversion. 

The mean and standard deviation of each distribution are displayed in Figure \ref{fig:fwi3d_misfit}b by black dots and red error bars, and are connected by a red line which we call a \textit{Bayesian L-curve}. From left to right along the curve, the magnitude of prior information (relative smoothness in this case) decreases from $\sigma_s = 31.25$ (top-left point) to $\sigma_s = \infty$ (no smoothing -- bottom-right point). An arrow marks the case with $\sigma_s = 1000$, and we observe that data misfits start to decrease significantly if we further increase the roughness value $\sigma_s$ of the prior information beyond this, implying that this value provides a balance between information from the smoothing distribution (used to narrow the possible parameter space) and the observed data (used to achieve good data fits). 

From a Bayesian point of view, if we do not know the smoothness of the true Earth, then we have little reason to choose any one of these posterior distributions -- they are all part of the posterior uncertainty. We may have sufficient prior intuition to suspect that posterior models are too rough in Figure \ref{fig:fwi3d_psvi_diff_smooth_vpr_mean_std_error}a and too smooth in Figure \ref{fig:fwi3d_psvi_diff_smooth_vpr_mean_std_error}h, but it is difficult to establish a probability distribution over levels of smoothness \textit{a priori}. Hierarchical Bayesian methods might be used to estimate this distribution \textit{a posteriori}. However, such an approach introduces physical inconsistency to Bayesian inversion results \cite{mosegaard2024inconsistency}. Alternatively, in a departure from Bayesian methods we might decide to entertain only a subset of these various results, for example those around $\sigma_s=1000$. 

More generally, many different types of prior hypotheses can be represented by a hypothesis space $\mathcal{H}$ represented schematically in Figure \ref{fig:fwi3d_misfit}c. Given a rather non-informative uniform prior distribution in the old posterior pdf, in principle we can inject any additional prior assumptions into the inversion; each would have different effects on the inversion results, between which we may wish to discriminate. As an example, we consider another prior assumption from real-geology. The geology informed prior information is obtained by subsampling a set of realistic geological images using a predefined window size. These subimages are used to calculate a local correlation matrix between parameter pairs that lie within that window size to define prior correlations between pairs of model parameters used in this test \cite{zhao2024variational}. We consider 5 such geological prior distributions constructed using different local cubic windows of edge length 2, 4, 6, 8 and 10 cells, respectively. Again, the corresponding posterior distributions are calculated using VPR by replacing the old uniform prior distribution by each geological prior distribution in turn. 

\begin{figure}
	\centering\includegraphics[width=\textwidth]{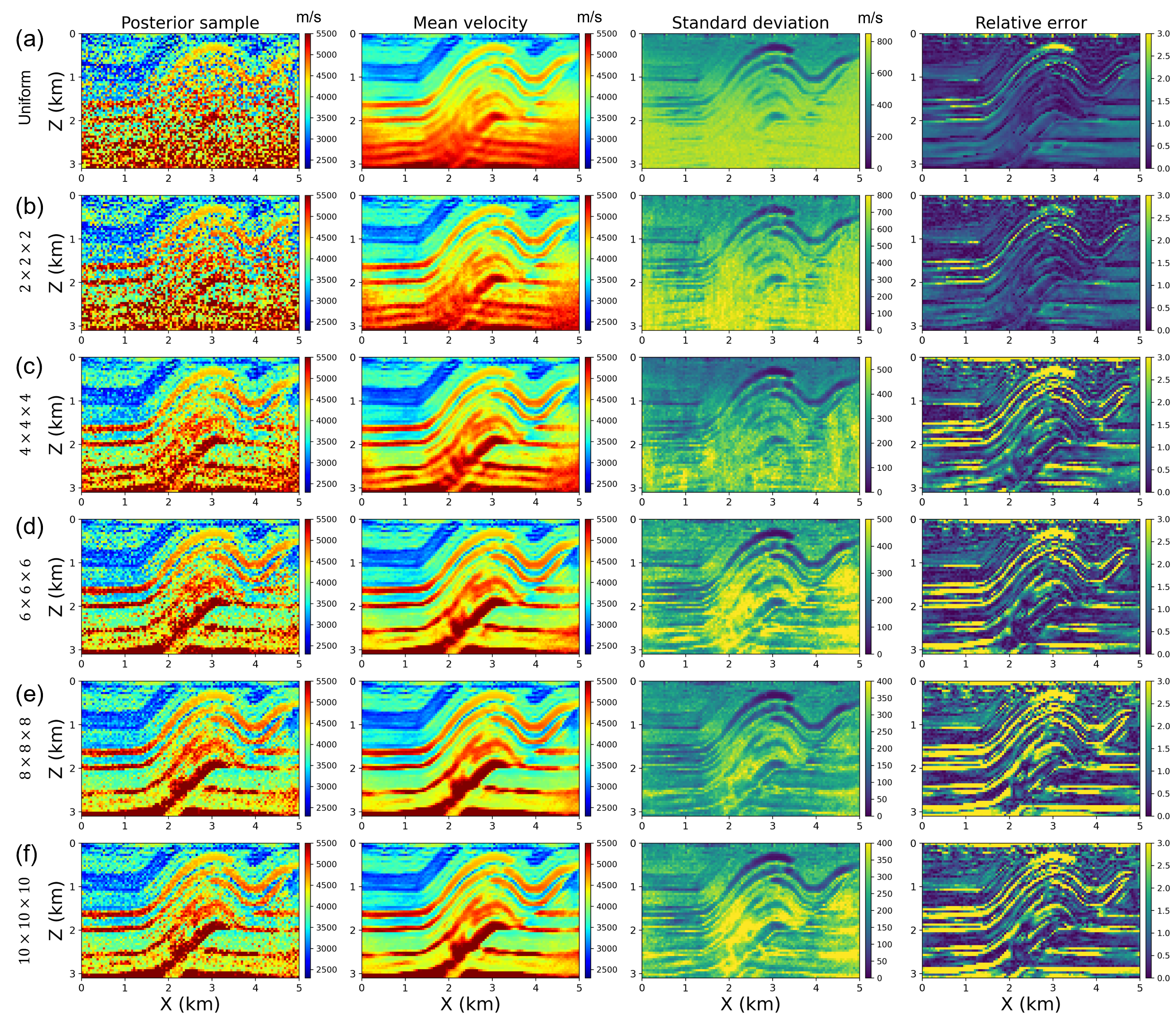}
	\caption{Inversion results on a vertical section Y = 2.5 km obtained using (a) the uniform prior pdf and (b) -- (f) the geological prior pdf's with different sizes of local correlation windows, denoted on the left side of each row. From left to right, they represent one posterior sample, the mean velocity, the standard deviation and the relative error maps.}
	\label{fig:fwi3d_psvi_diff_geological_vpr_mean_std_error_vertical_plot}
\end{figure}

Figure \ref{fig:fwi3d_psvi_diff_geological_vpr_mean_std_error_vertical_plot} displays the corresponding posterior pdf's, where the sizes of the correlation windows used are denoted on the left side of Figure \ref{fig:fwi3d_psvi_diff_geological_vpr_mean_std_error_vertical_plot}. Figure \ref{fig:fwi3d_psvi_diff_geological_vpr_mean_std_error_vertical_plot}a represents the results obtained using the uniform prior distribution. We observe that larger prior correlation windows generally result in better inversion results that are more similar to the true velocity model (Figure \ref{fig:fwi3d_true_vel_prior}c), since a larger window can define better prior correlation information, which improves the inversion results. However, note that a larger window also has higher memory requirements. 

These inversion results are then used to construct a Bayesian L-curve, as displayed by the dashed blue line in Figure \ref{fig:fwi3d_misfit}b. From left to right along this dashed blue line, the correlation window size decreases from $10\times10\times10$ (top-left point) to $0\times0\times0$ (bottom-right point), the latter being the uniform prior pdf with no prior correlation. This indicates that the strength of geological prior information decreases from left to right. The data misfits are lower than those from smoothing priors in red because geologically informed priors impose less smoothness vertically than horizontally (which is correct in the true model - and often in the Earth).


\subsection{Computational Cost}
The inversion using the uniform prior distribution (Figure \ref{fig:fwi3d_highf_3sections_uniform}) is performed within three frequency bands, each involving 200, 300 and 800 iterations for variational inference (optimisation) in the low, intermediate and high frequency bands respectively. In each iteration, 4 random samples are used to estimate expectations in the EBLO[$q(\mathbf{m})$] in equation \ref{eq:elbo}. Therefore, a total of 5200 samples, thus 5200 forward and adjoint simulations, are used to obtain the inversion results displayed in Figure \ref{fig:fwi3d_highf_3sections_uniform}. To further reduce the computational cost, we use a minibatch of 36 shot gathers, randomly selected from the total of 81 shots in each FWI simulation \cite{zhang20233}. For comparison, the same test conducted in \citet{zhang20233} were obtained using 4000, 400,000 and 80,000 samples (forward and adjoint simulations) when using automatic differentiation variational inference ADVI \cite[ADVI --][]{kucukelbir2017automatic}, Stein variational gradient descent \cite[SVGD --][]{liu2016stein}, and stochastic SVGD \cite[sSVGD --][]{gallego2018stochastic}, respectively. SVGD and sSVGD are far more expensive than PSVI; ADVI is computationally cheaper but provides biased inversion results with strongly underestimated posterior uncertainties \cite{zhang20233}. 

Note that in linearised, deterministic FWI, one would usually perform $\sim$100 iterations to obtain a reasonable result, and a full batch of 81 shots is normally used at each iteration to provide accurate gradient information. Since forward simulation using 81 shot gathers would require 2.25 times more computation than that using a minibatch of 36 shots, the computational cost of 100 iterations implemented in deterministic FWI is equivalent to 225 forward evaluations performed in this study.

With PSVI we therefore obtain reasonable probabilistic uncertainty estimates using only an order of magnitude more computation than deterministic FWI from which no robust uncertainty information can be obtained. In addition, this single set of variational inversion results allows multiple posterior distributions corresponding to different prior pdf's to be produced using VPR, without any further FWI simulations even in 3D.

\section{Discussion}
All of the posterior pdf's with different priors were obtained from only a single Bayesian 3D FWI using a uniform prior distribution. However, to translate uncertainties from model parameter space into data space, hundreds of additional forward simulations were performed to obtain data misfit values to calculate the Bayesian L-curves displayed in Figure \ref{fig:fwi3d_misfit}. Alternatively, \textit{boosting variational inference} (BVI) implicitly provides a small number (tens) of representative samples that represent major components of the uncertainties \cite{zhao2024bayesian}. Those representative samples might be used to construct Bayesian L-curves at a significantly reduced computational cost compared to the approach used here.

A common way to discriminate between prior hypotheses in Bayesian inference is to calculate the evidence term,
\begin{equation}
	p(\mathbf{d}_{obs})=\int_{\mathbf{m}}p(\mathbf{d}_{obs}|\mathbf{m})p(\mathbf{m})d\mathbf{m},
	\label{eq:evidence}
\end{equation}
for each hypothesis, which requires an integral over the entire model parameter space to be evaluated. While this is computationally possible for some low dimensional problems \cite{strutz2023variational}, for Bayesian 3D FWI it is infeasible due to the high dimensionality of the problem. Hence, the approach used here is the first to allow such models and prior hypotheses to be compared in a probabilistic manner and at feasible computational cost in 3D.

While we used prior distributions with varying degrees of smoothness to simulate different prior hypotheses, more realistic prior distributions that encode geological information can be employed in Bayesian inversion \cite{mosser2020stochastic, levy2022variational, bloem2022introducing, sun2024invertible}. By using the VPR framework we can calculate the corresponding posterior distributions and analyse different prior hypotheses at relatively low cost.

\begin{figure}
	\centering\includegraphics[width=\textwidth]{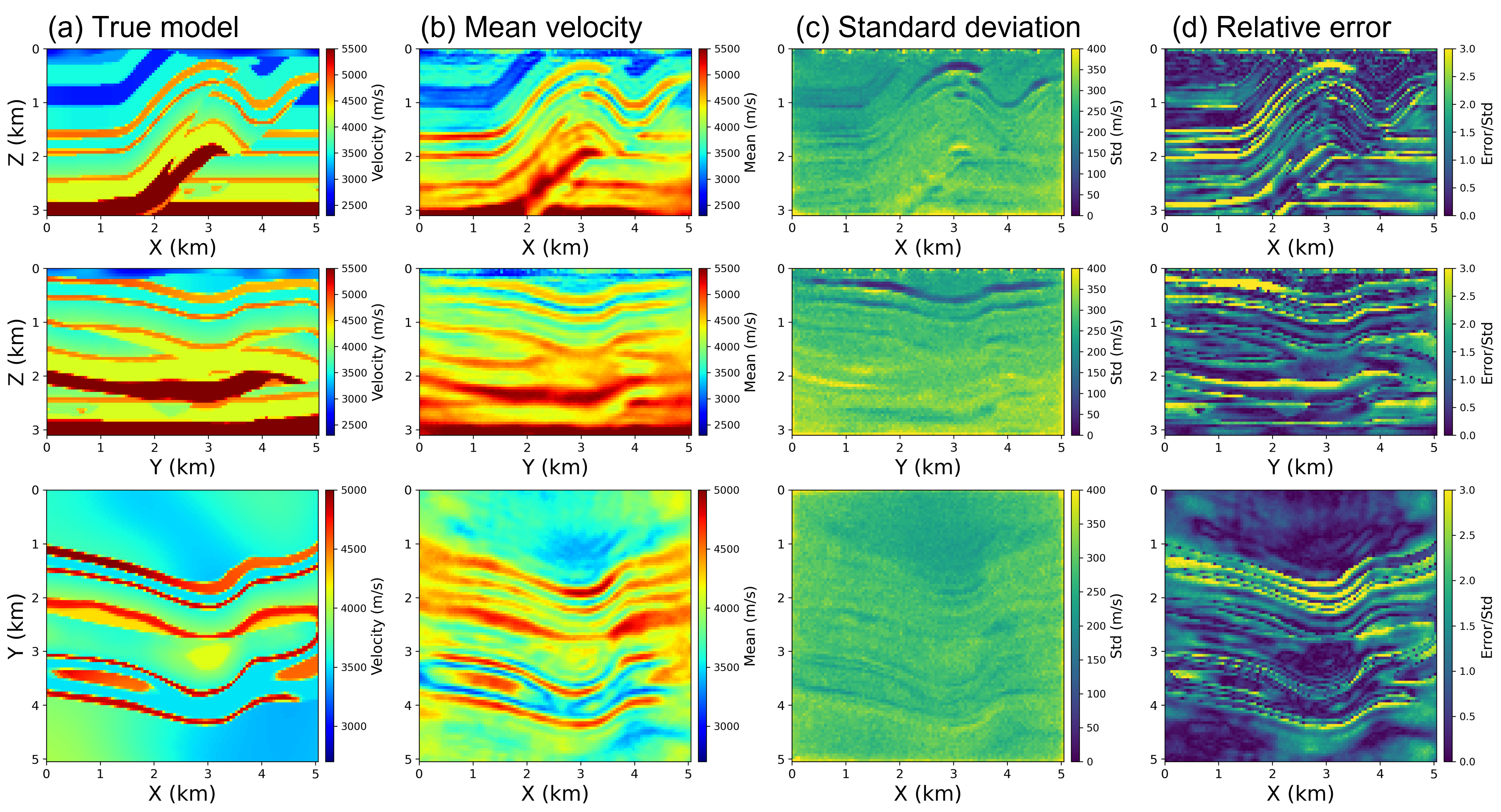}
	\caption{Inversion results after refining the VPR results, which are more similar to independent variational Bayesian inversion results (presented in Appendix B).}
	\label{fig:fwi3d_highf_3sections_smooth_vpr_with_finetune}
\end{figure}

Rather than approximating the true solution to the desired inverse problem directly, the VPR approximation to each new posterior pdf includes only information about the true solution that is already included within in the old posterior pdf. It therefore carries over an imprint of any errors in the latter \cite{zhao2024variational}. VPR results may therefore differ from independent Bayesian inversion results. This effect is potentially non-negligible due to the higher dimensionality and complexity of 3D Bayesian FWI problems as shown in Appendix B; qualitatively the results are less good than those obtained for 2D FWI in \citet{zhao2024variational}. To improve the results, we can use a relatively lower cost calculation to refine (fine tune) the outcomes obtained from VPR using observed data. We do this by invoking Bayes' rule again (performing another variational Bayesian inversion using PSVI), but with a smaller number of iterations which start from the VPR output because the VPR solution should already be reasonably close to the true posterior pdf. For example, we fine tuned the above VPR results by performing variational inversion with an additional 200 iterations using 4 samples per iteration. This costs an extra 800 forward and adjoint simulations, and the results are displayed in Figure \ref{fig:fwi3d_highf_3sections_smooth_vpr_with_finetune}. These are improved compared to the initial VPR results (Figure \ref{fig:fwi3d_highf_3sections_smooth_vpr}) and are more similar to independent inversion results (Figure \ref{fig:fwi3d_highf_3sections_smooth_psi}). If we wanted to reduce this additional computation further, we might again depart from formal Bayesian analysis by using the Bayesian L-curve obtained from VPR to select a smaller subset of prior hypotheses to entertain, then fine tune only those VPR posterior pdf's.

\section{Conclusions}
We solve a 3D Bayesian FWI problem using physically structured variational inference, in which a transformed Gaussian distribution is employed to approximate the posterior distribution of the fully nonlinear inverse problem. Using a non-informative uniform prior distribution we obtain a posterior solution with reasonable uncertainty estimates. The inversion progresses through low, intermediate and high frequency bands, with the total computational cost only an order of magnitude higher than deterministic FWI. In addition, we explore multiple different prior distributions and calculate the corresponding posterior distributions using variational prior replacement, without the need for additional FWI simulations. These results are used to analyse different prior hypotheses through the construction of Bayesian L-curves. We thus demonstrate that accurate Bayesian FWI and analysis or even a choice of prior hypotheses is in principle now feasible in 3D.


\section{ACKNOWLEDGMENTS}
We thank the Edinburgh Imaging Project (EIP - \url{https://blogs.ed.ac.uk/imaging/}) sponsors (BP and TotalEnergies) for supporting this research. We also thank Dr. Xin Zhang from China University of Geosciences in Beijing for providing insightful discussion about three variational inversion methods (ADVI, SVGD and sSVGD) and the corresponding inversion results.

\bibliographystyle{plainnat}  
\bibliography{reference}

\appendix

\section{Inversion results using low and intermediate frequency waveform data}
\label{ap:low_mid_freq}
\begin{figure}
	\centering\includegraphics[width=\textwidth]{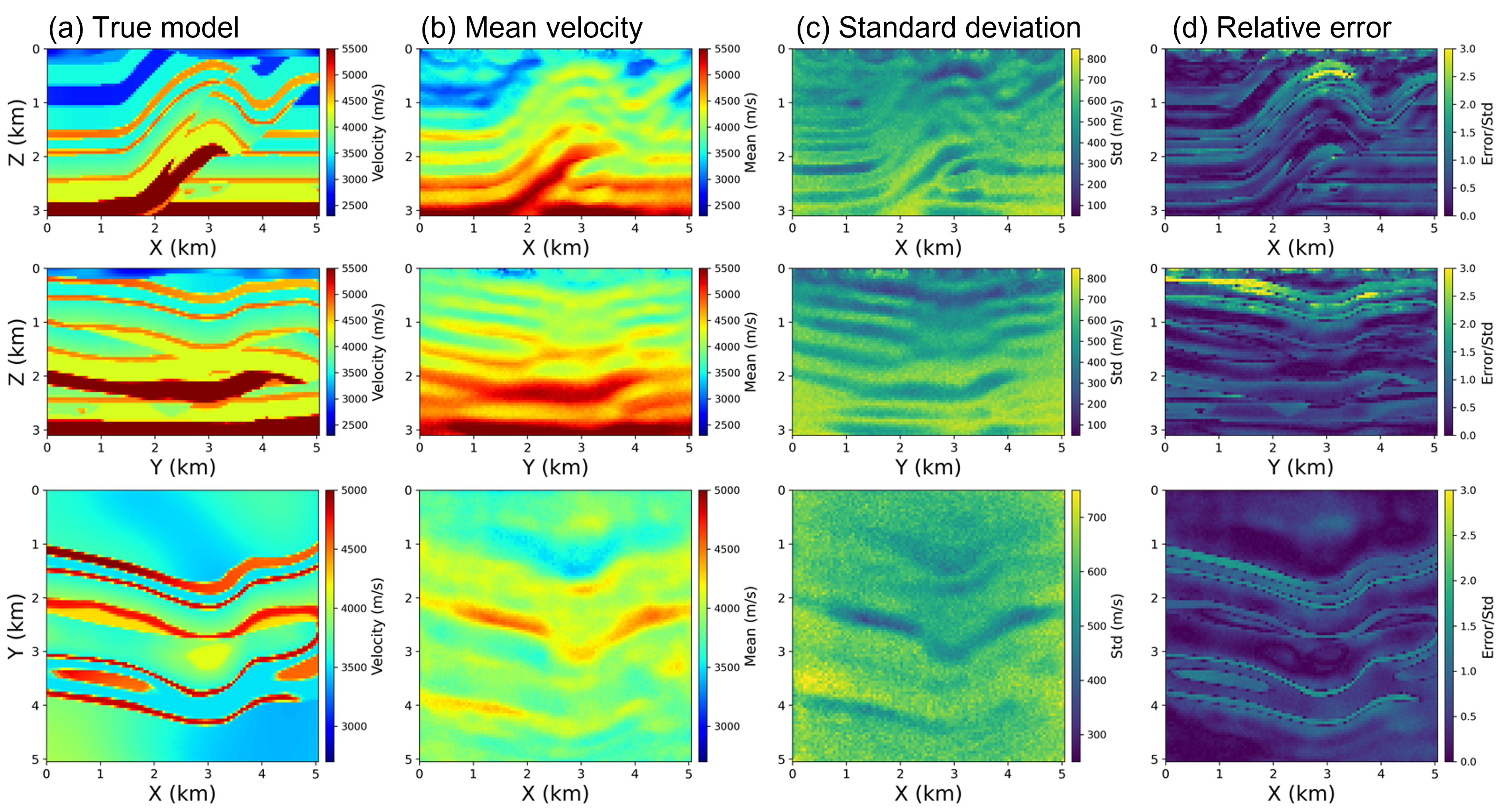}
	\caption{Low frequency 3D variational Bayesian FWI results obtained using PSVI and a uniform prior distribution. Top two rows show two vertical sections at Y = 2.5km and X = 2.5km, and bottom row shows a horizontal section at Z = 1.25km. Panels (a) -- (d) display the true velocity model, the mean, the standard deviation and the relative error maps of the posterior distribution.}
	\label{fig:fwi3d_lowf_3sections_uniform}
\end{figure}

\begin{figure}
	\centering\includegraphics[width=\textwidth]{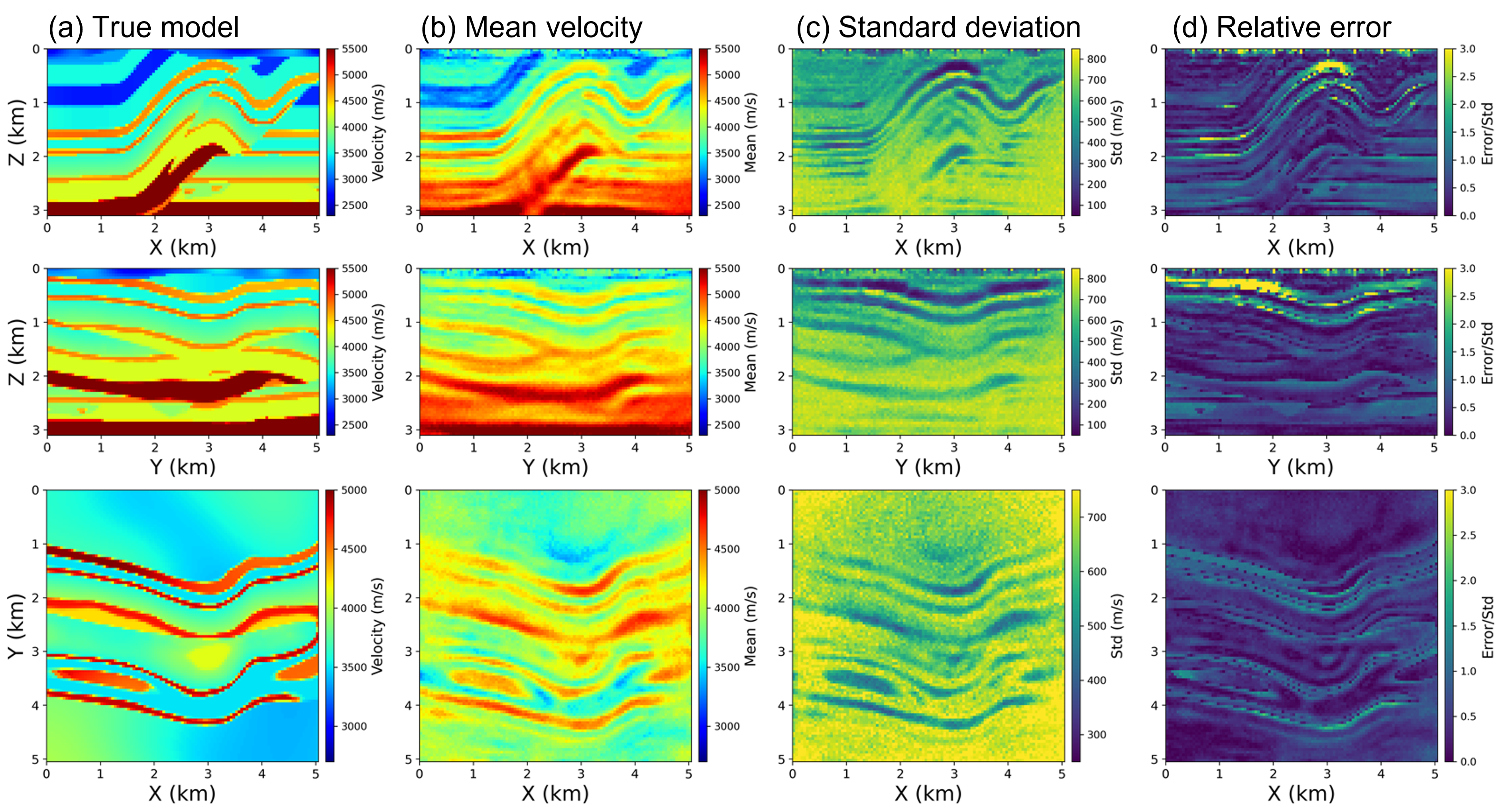}
	\caption{Intermediate frequency 3D variational Bayesian FWI results.}
	\label{fig:fwi3d_midf_3sections_uniform}
\end{figure}

Figures \ref{fig:fwi3d_lowf_3sections_uniform} and \ref{fig:fwi3d_midf_3sections_uniform}, respectively, display the inversion results obtained using low and intermediate frequency waveform data, performed using physically structured variational inference \cite[PSVI --][]{zhao2024physically}. Similarly to the inversion results displayed in Figure \ref{fig:fwi3d_highf_3sections_uniform} in the main text, the top two rows display two vertical sections at locations Y = 2.5 km and X = 2.5 km, and the bottom row illustrates one horizontal section at Z = 1.25 km. From left to right, each column shows three true velocity sections (marked by black dashed lines in Figure \ref{fig:fwi3d_true_vel_prior}a), the corresponding inverted average velocity, the standard deviation, and the relative error maps of the posterior distribution, where the relative error is the absolute difference between the true and mean values divided by the standard deviation at each point. Comparing these results with those using the high frequency data (Figure \ref{fig:fwi3d_highf_3sections_uniform}) we observe that as the frequency of waveform data used for inversion increases, the spatial resolution becomes higher, and velocity estimates become more accurate. The overall relative errors are within 3 standard deviations, which might be expected of the true posterior probability distribution.

\section{Verification of Variational Prior Replacement}
\label{ap:verify_vpr}

\begin{figure}
	\centering\includegraphics[width=\textwidth]{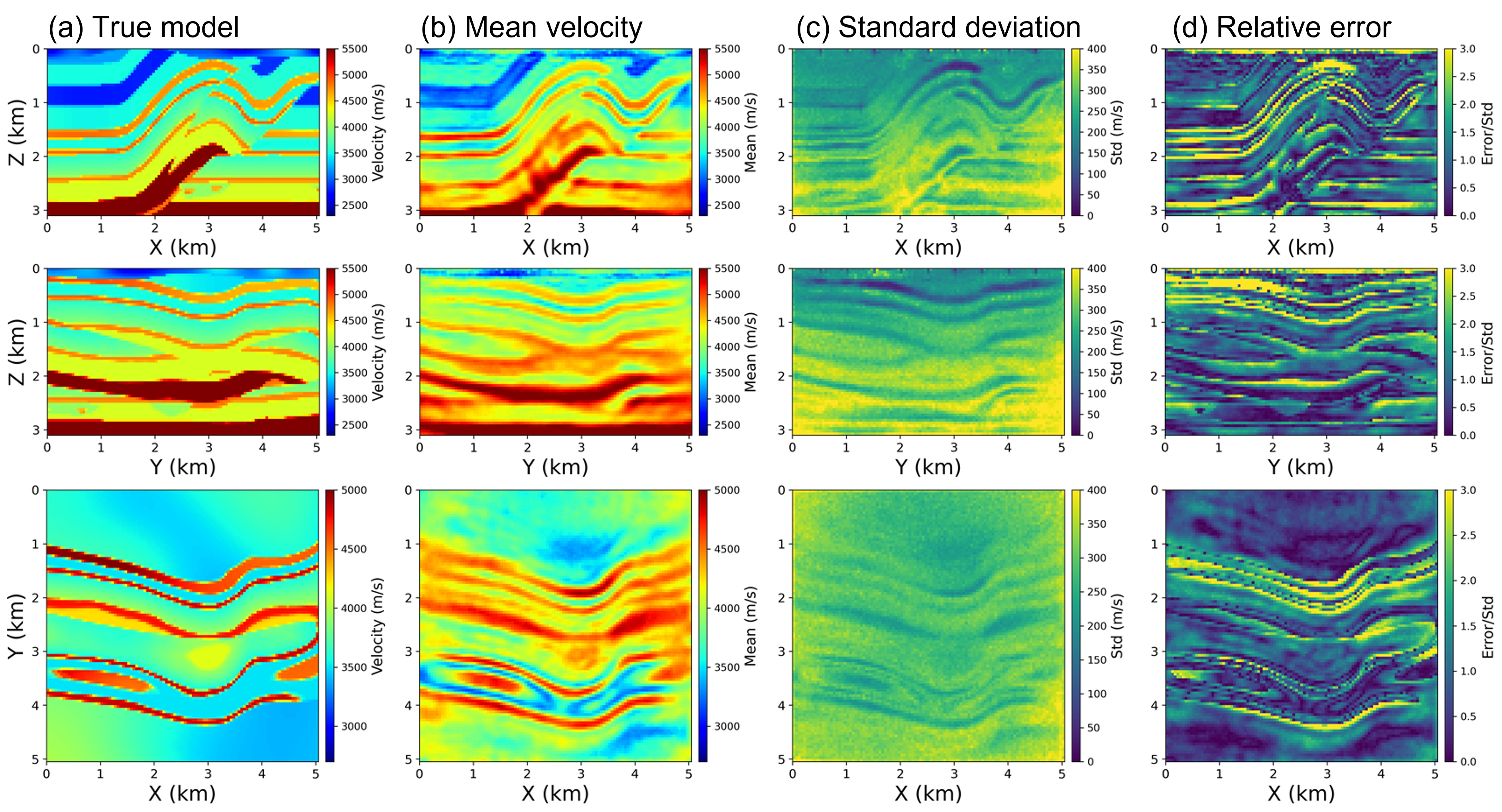}
	\caption{Variational prior replacement results obtained by removing the uniform prior information from the old posterior pdf, and imposing a smoothed prior pdf with a smoothness value of $\sigma_s = 1000$.}
	\label{fig:fwi3d_highf_3sections_smooth_vpr}
\end{figure}

\begin{figure}
	\centering\includegraphics[width=\textwidth]{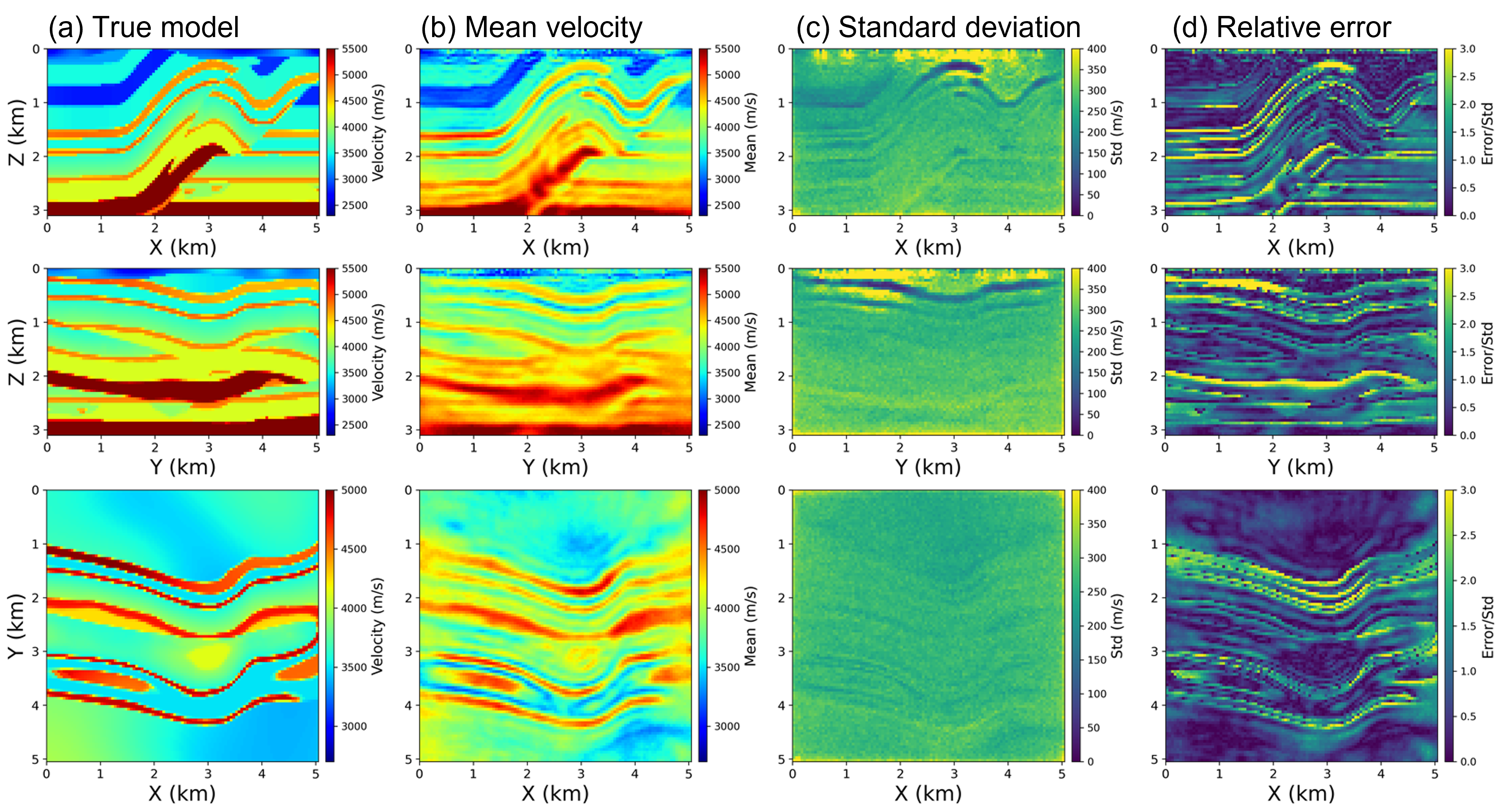}
	\caption{Independent variational Bayesian inversion results obtained using the smoothed prior distribution with a smoothness value of $\sigma_s = 1000$. This is used to verify VPR results displayed in Figure \ref{fig:fwi3d_highf_3sections_smooth_vpr}.}
	\label{fig:fwi3d_highf_3sections_smooth_psi}
\end{figure}

In this Appendix, we test the effectiveness and accuracy of variational prior replacement \cite[VPR --][]{zhao2024variational} for 3D Bayesian FWI by comparing the posterior solutions obtained using VPR to those obtained from independent Bayesian inversions. For VPR, we consider the uniform prior distribution defined in the main text as the old prior pdf, which is then removed from the old posterior distribution (displayed in Figure \ref{fig:fwi3d_highf_3sections_uniform} in the main text) and replaced by a smoothed prior distribution with $\sigma_s = 1000$ (see details in the main text). The corresponding results are displayed in Figure \ref{fig:fwi3d_highf_3sections_smooth_vpr}. We then perform an independent variational Bayesian inversion using the same smoothed prior distribution. Similarly to the uniform prior case, the inversion is performed using the same three frequency bands of waveform data with the same number of forward simulations. Figure \ref{fig:fwi3d_highf_3sections_smooth_psi} illustrates the results. The main features of these two results are roughly consistent, indicating that VPR can replace the old with the new prior information in the inversion results without solving a Bayesian inverse problem repeatedly from scratch. Some small discrepancies between these two sets of results remain, especially in the deeper parts of the model. This is partly because VPR introduces a variational distribution $q_{new}(\mathbf{m})$ to approximate the new posterior pdf $p_{new}(\mathbf{m}|\mathbf{d}_{obs})$, rather than calculating the true new posterior distribution, as expressed in the fifth line in equation \ref{eq:bayes_vpr2} in the main text \citep{walker2014varying, zhao2024variational}. In addition, this also indicates that VPR might not be as accurate for this extremely high dimensional and complex 3D Bayesian inverse problem, compared to previous tests in 2D \citep{zhao2024variational}. Nevertheless, this result is obtained almost for free, compared to the huge computational cost typically involved in 3D forward simulation \citep{zhao2020domain} and Bayesian inversion \citep{zhang20233}. In the main text, we introduce a fine tuning scheme to improve the inversion results displayed in Figure \ref{fig:fwi3d_highf_3sections_smooth_vpr}.

\label{lastpage}
\end{document}